\documentclass[dvips]{article}
\usepackage{amssymb}
\usepackage{amsmath}
\usepackage{rotating}
\DeclareSymbolFont{ppa}{OT1}{ppl}{m}{it}
\DeclareMathSymbol{\vv}{\mathalpha}{ppa}{'166}

\begin{document}

\newcommand{\dd}{\,{\rm d}}
\newcommand{\ie}{{\it i.e.},\,}
\newcommand{\etal}{{\it et al.\ }}
\newcommand{\eg}{{\it e.g.},\,}
\newcommand{\cf}{{\it cf.\ }}
\newcommand{\vs}{{\it vs.\ }}
\newcommand{\zdot}{\makebox[0pt][l]{.}}
\newcommand{\up}[1]{\ifmmode^{\rm #1}\else$^{\rm #1}$\fi}
\newcommand{\dn}[1]{\ifmmode_{\rm #1}\else$_{\rm #1}$\fi}
\newcommand{\upd}{\up{d}}
\newcommand{\uph}{\up{h}}
\newcommand{\upm}{\up{m}}
\newcommand{\ups}{\up{s}}
\newcommand{\arcd}{\ifmmode^{\circ}\else$^{\circ}$\fi}
\newcommand{\arcm}{\ifmmode{'}\else$'$\fi}
\newcommand{\arcs}{\ifmmode{''}\else$''$\fi}
\newcommand{\MS}{{\rm M}\ifmmode_{\odot}\else$_{\odot}$\fi}
\newcommand{\RS}{{\rm R}\ifmmode_{\odot}\else$_{\odot}$\fi}
\newcommand{\LS}{{\rm L}\ifmmode_{\odot}\else$_{\odot}$\fi}
\newcommand{\feh}{\hbox{$ [{\rm Fe}/{\rm H}]$}}

\newcommand{\Abstract}[2]{{\footnotesize\begin{center}ABSTRACT\end{center}
\vspace{1mm}\par#1\par
\noindent
{~}{\it #2}}}

\newcommand{\TabCap}[2]{\begin{center}\parbox[t]{#1}{\begin{center}
  \small {\spaceskip 2pt plus 1pt minus 1pt T a b l e}
  \refstepcounter{table}\thetable \\[2mm]
  \footnotesize #2 \end{center}}\end{center}}

\newcommand{\TableSep}[2]{\begin{table}[p]\vspace{#1}
\TabCap{#2}\end{table}}

\newcommand{\FigCap}[1]{\footnotesize\par\noindent Fig.\  %
  \refstepcounter{figure}\thefigure. #1\par}

\newcommand{\TableFont}{\footnotesize}
\newcommand{\TableFontIt}{\ttit}
\newcommand{\SetTableFont}[1]{\renewcommand{\TableFont}{#1}}

\newcommand{\MakeTable}[4]{\begin{table}[htb]\TabCap{#2}{#3}
  \begin{center} \TableFont \begin{tabular}{#1} #4
  \end{tabular}\end{center}\end{table}}

\newcommand{\MakeTableSep}[4]{\begin{table}[p]\TabCap{#2}{#3}
  \begin{center} \TableFont \begin{tabular}{#1} #4
  \end{tabular}\end{center}\end{table}}

\newenvironment{references}%
{
\footnotesize \frenchspacing
\renewcommand{\thesection}{}
\renewcommand{\in}{{\rm in }}
\renewcommand{\AA}{Astron.\ Astrophys.}
\newcommand{\AAS}{Astron.~Astrophys.~Suppl.~Ser.}
\newcommand{\ApJ}{Astrophys.\ J.}
\newcommand{\ApJS}{Astrophys.\ J.~Suppl.~Ser.}
\newcommand{\ApJL}{Astrophys.\ J.~Letters}
\newcommand{\AJ}{Astron.\ J.}
\newcommand{\IBVS}{IBVS}
\newcommand{\PASJ}{PASJ}
\newcommand{\PASP}{P.A.S.P.}
\newcommand{\Acta}{Acta Astron.}
\newcommand{\MNRAS}{MNRAS}
\renewcommand{\and}{{\rm and }}
\section{{\rm REFERENCES}}
\sloppy \hyphenpenalty10000
\begin{list}{}{\leftmargin1cm\listparindent-1cm
\itemindent\listparindent\parsep0pt\itemsep0pt}}%
{\end{list}\vspace{2mm}}

\def\TYLDA{~}
\newlength{\DW}
\settowidth{\DW}{0}
\newcommand{\dw}{\hspace{\DW}}

\newcommand{\refitem}[5]{\item[]{#1} #2%
\def\REFARG{#3}\ifx\REFARG\TYLDA\else, {\it#3}\fi
\def\REFARG{#4}\ifx\REFARG\TYLDA\else, {\bf#4}\fi
\def\REFARG{#5}\ifx\REFARG\TYLDA\else, {#5}\fi.}

\newcommand{\Section}[1]{\section{#1}}
\newcommand{\Subsection}[1]{\subsection{#1}}
\newcommand{\Acknow}[1]{\par\vspace{5mm}{\bf Acknowledgments.} #1}
\pagestyle{myheadings}

\newfont{\bb}{ptmbi8t at 12pt}
\newcommand{\xrule}{\rule{0pt}{2.5ex}}
\newcommand{\xxrule}{\rule[-1.8ex]{0pt}{4.5ex}}
\def\thefootnote{\fnsymbol{footnote}}

\begin{center}
{\Large\bf A Low-Resolution Spectroscopic Exploration\\
of Puzzling OGLE Variable Stars\footnote{Based on observations
obtained with the 2.5-m Ir\'en\'ee du Pont telescope at the Las Campanas
Observatory of the Carnegie Institution for Science under
the CNTAC program CN2014A-65.}}
\vskip1cm
{\bf
P.~~P~i~e~t~r~u~k~o~w~i~c~z$^1$,~~M.~~L~a~t~o~u~r$^2$,~~R.~~A~n~g~e~l~o~n~i$^3$,\\
F.~~d~i~~M~i~l~l~e$^4$,~~I.~~S~o~s~z~y~\'n~s~k~i$^1$,~~A.~U~d~a~l~s~k~i$^1$,\\
~~and~~C.~G~e~r~m~a~n~\`a$^5$\\}
\vskip3mm
{
$^1$ Warsaw University Observatory, Al. Ujazdowskie 4, 00-478 Warszawa, Poland\\
e-mail: pietruk@astrouw.edu.pl\\
$^2$ Dr. Karl Remeis-Observatory \& ECAP, Astronomical Institute,\\
Friedrich-Alexander University Erlangen-Nuremberg,\\
Sternwartstrasse 7, D-96049 Bamberg, Germany\\
$^3$ Gemini Observatory, Casilla 603, La Serena, Chile\\
$^4$ Las Campanas Observatory, Casilla 601, La Serena, Chile\\
$^5$ Departamento de F\'isica, Universidade Federal do Maranh\~ao,
S\~ao Lu\'is, MA, Brazil\\
}
\end{center}

\Abstract{We present the results of a spectroscopic follow-up of various
puzzling variable objects detected in the OGLE-III Galactic disk and bulge
fields. The sample includes mainly short-period multi-mode pulsating
stars that could not have been unambiguously classified as either $\delta$ Sct
or $\beta$ Cep type stars based on photometric data only, also stars with
irregular fluctuations mimicking cataclysmic variables and stars with
dusty shells, and periodic variables displaying brightenings in their light curves
that last for more than half of the period. The obtained low-resolution
spectra show that all observed short-period pulsators are of $\delta$ Sct
type, the stars with irregular fluctuations are young stellar objects,
and the objects with regular brightenings are A type stars or very likely
Ap stars with strong magnetic field responsible for the presence
of bright caps around magnetic poles on their surface. We also took spectra
of objects designated OGLE-GD-DSCT-0058 and OGLE-GD-CEP-0013. An estimated
effective temperature of 33~000~K in OGLE-GD-DSCT-0058 indicates that it
cannot be a $\delta$ Sct type variable. This very short-period (0.01962~d)
high-amplitude (0.24~mag in the $I$-band) object remains a mystery.
It may represent a new class of variable stars. The spectrum of
OGLE-GD-CEP-0013 confirms that this is a classical Cepheid despite a
peculiar shape of its light curve. The presented results will help in proper
classification of variable objects in the OGLE Galaxy Variability Survey.}

{Stars: variables: Cepheids -- Stars: variables: delta
Scuti -- Stars: variables: T Tauri, Herbig Ae/Be -- Stars: peculiar}


\Section{Introduction}

The Optical Gravitational Lensing Experiment (OGLE) is a long-term
large-scale sky variability survey conducted at Las Campanas Observatory,
Chile. It started in 1992 with the prime aim of searches for microlensing
events in dense stellar regions of the Galactic bulge (Udalski \etal 1992).
In 1997 the project entered the second phase (OGLE-II) with the beginning
of operation of the dedicated 1.3-m Warsaw Telescope. An upgrade from a
single-chip camera to a 8-chip mosaic camera started its third phase (OGLE-III)
in 2001, which lasted until 2009 (Udalski 2003). Since March 2010 the
project is in its fourth phase. The OGLE-IV camera consists of 32 chips
and has a field of view of about 1.4~deg$^2$. Currently, OGLE measures
brightness in the $I$ and $V$ bands of over a billion stars of the
Galactic bulge and disk and the Magellanic System, covering a total area
of about 3000 deg$^2$ of the sky (see Udalski, Szyma\'nski and Szyma\'nski
2015 for details).

Regular observations conducted for many years allow finding and exploring the
variety of variable objects (\eg Soszy\'nski \etal 2008, 2013, 2014, Poleski
\etal 2010, Mr\'oz \etal 2013, Pietrukowicz \etal 2013a). However, for some
variables the obtained photometry is insufficient to properly classify the
objects or even explain their nature. It happens sometimes in the case of Milky Way
stars with unknown distance and reddening. For this reason we have conducted
a spectroscopic follow-up of selected puzzling objects detected in the
OGLE-III fields toward the Galactic disk and bulge. Among the targets are
pulsating stars that could not have been properly classified based
on photometry only (Pietrukowicz \etal 2013b). Another group is formed of
variables showing irregular fluctuations up to 2.0~mag in their light curves.
We also took spectra of several periodic variables (periods of a few days)
that exhibit brightenings or waves lasting for more than half of the period.


\Section{Spectroscopic Observations and Reductions}

The spectra were obtained with the 2.5-m Ir\'en\'ee du Pont
telescope at Las Campanas Observatory on two nights,
April 27/28 and April 28/29, 2014. The observatory is operated
by the Carnegie Institution for Science. We used the Boller
and Chivens spectrograph with the lowest grating of 300 line/mm
giving a resolution of 3.0 \AA/pixel with a wavelength coverage
of about 6200 \AA. The spatial scale on the du Pont telescope is 0.70
arcsec/pixel and the slit length 271$\arcs$. The two observing nights
were generally clear with some cirrus clouds at the end of the second one.
For the first seven hours of the night we took spectra of stars
in the OGLE-III Galactic disk fields while for the last three hours
selected stars in the OGLE-III bulge area. Since all targets are
relatively faint objects located in dense fields, special care was taken
to precise pointing of the instrument and aligning the slit.
For the majority of observations the slit was aligned at the parallactic
angle to avoid light loss due differential refraction, but in several
cases we had to set a different angle to avoid contamination from neighboring
stars. The observed variable stars are listed in Table~1, in which we give
information on coordinates, average brightness, exposure times,
and the reason of taking the spectrum. For each program star we took two or
three exposures and one 90-s Ne-He-Ar lamp exposure in between
for the wavelength calibrations. Additional calibration images included
bias and dome flat-field images, and spectra of flux standards.

\begin{sidewaystable}[htb!]
\centering \caption{\small OGLE variable stars for which spectra were
obtained}
\medskip
{\scriptsize
\begin{tabular}{lcccccccccc}
\hline
ID or name        &      RA(2000)     &      Dec(2000)     & $\langle V \rangle$ & $\langle I \rangle$ & $A_{1,I}$ & $P_1$ & $P_2$ & $P_3$ & $t_{\rm exp}$ & Puzzle\\
                  &                   &                    &        [mag]        &        [mag]        &   [mag]   &  [d]  &  [d]  &  [d]  &      [s]      & \\
\hline
CAR118.5.9107     & $10\uph36\upm24\zdot\ups80$ & $-63\arcd09\arcm03\zdot\arcs3$ & 16.66 & 15.90 & 0.115 & 0.12655899(9)  & 0.1229418(5)  & 0.0735669(4)$^*$   & $3\times600$  & $\delta$ Sct or $\beta$ Cep? \\
CAR116.5.7705     & $10\uph36\upm27\zdot\ups08$ & $-62\arcd35\arcm08\zdot\arcs1$ & 15.93 & 15.05 & 0.048 & 0.1192127(2)   & 0.1670429(12) & 0.198133(2)$^*$    & $2\times750$  & $\delta$ Sct or $\beta$ Cep? \\
CAR116.2.30884    & $10\uph37\upm43\zdot\ups31$ & $-62\arcd46\arcm50\zdot\arcs9$ & 15.08 & 14.55 & 0.046 & 0.05203784(5)  & 0.05318510(7) & 0.05106081(5)      & $2\times480$  & $\delta$ Sct or $\beta$ Cep? \\
CAR115.6.12025    & $10\uph38\upm51\zdot\ups55$ & $-62\arcd04\arcm49\zdot\arcs6$ & 14.25 & 13.56 & 0.049 & 0.1854823(8)   & 0.1847128(6)  & 0.1812189(4)$^*$   & $2\times180$  & $\delta$ Sct or $\beta$ Cep? \\ 
CAR118.4.1153     & $10\uph39\upm00\zdot\ups95$ & $-63\arcd11\arcm04\zdot\arcs0$ & 15.11 & 14.27 & 0.043 & 0.1670464(4)   & 0.1595196(8)  & 0.1709402(8)$^*$   & $2\times480$  & $\delta$ Sct or $\beta$ Cep? \\
CAR110.7.13663    & $10\uph40\upm31\zdot\ups51$ & $-61\arcd39\arcm58\zdot\arcs0$ & 14.67 & 13.90 & 0.032 & 0.1053973(5)   & 0.1164492(10) & 0.1266376(13)$^*$  & $2\times300$  & $\delta$ Sct or $\beta$ Cep? \\
CEN106.6.8162     & $11\uph31\upm35\zdot\ups92$ & $-60\arcd47\arcm45\zdot\arcs1$ & 14.94 & 14.13 & 0.025 & 0.140374(11)   & 0.135072(7)   & 0.14592(3)$^*$     & $3\times150$  & $\delta$ Sct or $\beta$ Cep? \\
CEN106.2.38514    & $11\uph33\upm28\zdot\ups13$ & $-60\arcd53\arcm44\zdot\arcs1$ & 15.80 & 14.80 & 0.036 & 0.133966(8)    & 0.138426(9)   & 0.130682(10)       & $3\times300$  & $\delta$ Sct or $\beta$ Cep? \\
CEN107.2.26501    & $11\uph55\upm39\zdot\ups22$ & $-62\arcd05\arcm11\zdot\arcs6$ & 14.87 & 14.09 & 0.028 & 0.116041(7)    & 0.0624465(12) & 0.12649(3)$^*$     & $3\times300$  & $\delta$ Sct or $\beta$ Cep? \\
CEN107.4.32037    & $11\uph56\upm12\zdot\ups94$ & $-61\arcd47\arcm44\zdot\arcs7$ & 14.96 & 13.96 & 0.037 & 0.104286(6)    & 0.107450(5)   & 0.169142(17)$^*$   & $3\times300$  & $\delta$ Sct or $\beta$ Cep? \\
CEN108.6.93566    & $13\uph31\upm32\zdot\ups10$ & $-64\arcd06\arcm46\zdot\arcs4$ & 16.28 & 15.05 & 0.021 & 0.08542011(4)  & 0.2138035(6)  & 0.08542535(15)     & $2\times900$  & $\delta$ Sct or $\beta$ Cep? \\
CEN108.4.43244    & $13\uph34\upm43\zdot\ups73$ & $-64\arcd02\arcm48\zdot\arcs1$ & 16.30 & 15.06 & 0.074 & 0.12975290(2)  & 0.1648194(2)  & 0.14203803(16)$^*$ & $2\times900$  & $\delta$ Sct or $\beta$ Cep? \\
MUS101.3.34906    & $13\uph26\upm16\zdot\ups46$ & $-64\arcd55\arcm40\zdot\arcs0$ & 16.71 & 15.60 & 0.104 & 0.14610250(1)  & 0.11689411(5) & 0.09741818(13)$^*$ & $2\times900$  & $\delta$ Sct? \\
OGLE-GD-DSCT-0012 & $10\uph42\upm51\zdot\ups62$ & $-61\arcd35\arcm21\zdot\arcs4$ & 14.17 & 13.69 & 0.060 & 0.1674796(13)  & 0.1315341(11) & 0.1323302(9)$^*$   & $3\times120$  & $\delta$ Sct? \\
OGLE-GD-DSCT-0058 & $10\uph41\upm48\zdot\ups77$ & $-61\arcd25\arcm08\zdot\arcs5$ & 17.71 & 17.22 & 0.231 & 0.01962154(1)  & 0.01891531(4) & 0.02038256(6)      & $3\times600$  & $\delta$ Sct? \\
OGLE-GD-CEP-0013  & $11\uph33\upm02\zdot\ups68$ & $-60\arcd52\arcm04\zdot\arcs5$ & 16.73 & 15.11 & 0.293 & 5.2436(9)      &       -       &        -           & $3\times600$  & $\delta$ Cep? \\
\hline
CAR117.5.6157     & $10\uph41\upm12\zdot\ups18$ & $-62\arcd33\arcm42\zdot\arcs6$ & 18.05 & 16.75 & 1.35  &       -        &       -       &        -           & $2\times1800$ & CV or dusty? \\
MUS100.3.59946    & $13\uph16\upm20\zdot\ups52$ & $-64\arcd44\arcm55\zdot\arcs0$ & 16.27 & 15.25 & 0.48  &       -        &       -       &        -           & $2\times900$  & dusty? \\
BLG179.1.111115   & $17\uph50\upm29\zdot\ups48$ & $-30\arcd50\arcm18\zdot\arcs1$ & 18.17 & 15.26 & 2.10  &       -        &       -       &        -           & $2\times1800$ & dusty? \\
\hline
CAR106.7.14162    & $11\uph00\upm40\zdot\ups09$ & $-61\arcd54\arcm49\zdot\arcs3$ & 15.08 & 14.38 & 0.14  & 3.20182(6)     &       -       &        -           & $2\times900$  & magnetic? \\
CAR106.7.46218    & $11\uph01\upm26\zdot\ups94$ & $-61\arcd51\arcm02\zdot\arcs2$ & 15.61 & 15.12 & 0.207 & 5.21579(3)     &       -       &        -           & $2\times900$  & magnetic? \\
BLG130.2.137067   & $17\uph47\upm44\zdot\ups92$ & $-34\arcd16\arcm15\zdot\arcs2$ & 13.72 & 13.01 & 0.063 & 6.59504(9)     &       -       &        -           & $2\times600$  & magnetic? \\
BLG183.4.156561   & $18\uph01\upm43\zdot\ups65$ & $-30\arcd24\arcm54\zdot\arcs0$ & 14.36 & 13.70 & 0.055 & 1.524444(5)    &       -       &        -           & $2\times600$  & magnetic? \\
BLG249.1.133775   & $18\uph06\upm07\zdot\ups50$ & $-26\arcd05\arcm08\zdot\arcs5$ & 13.91 & 13.08 & 0.075 & 3.77762(4)     &       -       &        -           & $2\times600$  & magnetic? \\
\hline
OGLE-GD-DSCT-0011 & $10\uph42\upm45\zdot\ups67$ & $-61\arcd35\arcm37\zdot\arcs2$ & 14.60 & 13.93 & 0.303 & 0.11309448(3)  &       -       &        -           & $3\times100$  & $\delta$ Sct \\
ASAS111308-6106.8 & $11\uph13\upm07\zdot\ups29$ & $-61\arcd06\arcm50\zdot\arcs0$ &  9.53 &  9.20 & 0.033 & 0.1931041(15)  &       -       &        -           & $2\times30$   & $\beta$ Cep \\
\hline
\end{tabular}
\\The group of candidate pulsating variables is followed by
a group of variables with irregular light variations and periodic
variables with waves in their light curves. The last two objects are
{\it bona fide} $\delta$ Sct and $\beta$ Cep type pulsators which
spectra served for comparison. Asterisks in the column with $P_3$
denote stars in which more than three modes were detected
(Pietrukowicz \etal 2013b).}
\end{sidewaystable}

For the reductions we used the utilities provided in the IRAF package
\footnote{IRAF is distributed by the National Optical Astronomy Observatory,
which is operated by the Association of Universities for Research
in Astronomy, Inc., under a cooperative agreement with the National
Science Foundation.}. In the first step, we combined the bias and flat-field
frames and then processed the science images through trimming, bias
and flat-field correction. In the next step, we extracted one-dimentional
spectra and applied the wavelength calibration to all single spectra.
By combining multiple observations of the same objects we improved
signal to noise and removed deviant pixels caused by cosmic rays. 
Based on the spectra of flux standards we determined the sensitivity
and extinction functions and applied them to all program objects.
Finally, in normalized spectra we measured equivalent widths of selected
lines and determined spectral types of our objects (based on Jaschek and
Jaschek 1987). They are compiled in Table~2.

\begin{table}[htb!]
\centering \caption{\small Equivalent widths (in \AA)~of some lines in the spectra
of puzzling variables}
\medskip
{\small
\begin{tabular}{lcccccc}
\hline
ID or name        & Ca{\scriptsize II}K & Ca{\scriptsize II}H+H$\epsilon$ & H$\gamma$ & H$\delta$ & G band & Sp. type \\
                  & 3934 & 3968 & 4102 & 4340 & 4307 & \\
\hline
CAR118.5.9107     &  5.7  &  7.8  &  6.4  &  5.6  &       & F0 \\
CAR116.5.7705     &  6.5  &  7.6  &  5.4  &  5.7  &       & F0 \\
CAR116.2.30884    &  3.6  &  8.9  &  9.5  & 10.1  &       & A5 \\
CAR115.6.12025    &  6.1  &  8.6  &  7.1  &  8.1  &       & F0 \\
CAR118.4.1153     &  5.9  &  7.9  &  5.7  &  5.7  &       & F0 \\
CAR110.7.13663    &  6.4  &  8.2  &  6.1  &  6.8  &       & F0 \\
CEN106.6.8162     &  5.2  &  8.2  &  5.3  &  7.4  &       & F0 \\
CEN106.2.38514    &  3.6  &  6.7  &  4.6  &  5.7  &       & A7 \\
CEN107.2.26501    &  5.7  &  8.4  &  7.0  &  8.8  &       & F0 \\
CEN107.4.32037    &  7.7  &  8.6  &  6.6  &  6.6  &       & F2 \\
CEN108.6.93566    &  5.0  &  8.7  &  6.3  &  7.5  &       & F0 \\
CEN108.4.43244    &  6.0  &  9.7  &  9.2  & 10.0  &       & F0 \\
MUS101.3.34906    &  3.5  &  7.1  &  7.0  &  7.6  &       & A7 \\
OGLE-GD-DSCT-0012 &  5.4  &  8.5  &  7.1  &  7.3  &       & F0 \\
OGLE-GD-DSCT-0058 &       &       &  3.4  &  4.2  &       & O9 \\
OGLE-GD-CEP-0013  & 16.1  & 10.7  &       &  0.8  &  7.8  & G0 \\
\hline
CAR117.5.6157     &  7.0  &  2.0  &       &       &  3.2  & G3 \\
MUS100.3.59946    &  6.4  &  7.1  &  4.6  &  4.6  &  3.8  & F5 \\
BLG179.1.111115   &  1.9  &  9.8  &  9.1  & 12.4  &       & A5 \\
\hline
CAR106.7.14162    &  4.8  &  7.6  &  8.4  &  7.1  &       & A7 \\
CAR106.7.46218    &  1.3  & 10.2  & 11.1  & 11.1  &       & A0 \\
BLG130.2.137067   &  0.4  &  7.7  &  7.6  &  7.8  &       & A0 \\
BLG183.4.156561   &  0.7  &  9.1  &  8.6  &  9.4  &       & A0 \\
BLG249.1.133775   &  0.6  &  9.9  & 10.3  & 10.5  &       & A0 \\
\hline
\end{tabular}}
\end{table}


\Section{Results}

\Subsection{Pulsating Stars}

Pietrukowicz \etal (2013b) reported the identification of 221 pulsating
stars and candidates for such objects in the OGLE-III Galactic disk fields.
Sixty of the stars are short-period multi-mode pulsators that are either
of $\delta$ Sct or $\beta$ Cep type. The main mode in these stars has
a period $0.05<P_{\rm main}<0.19$~d and $I$-band amplitude between
$0.025$~mag and $0.12$~mag. Due to unknown distance and reddening the
objects could not be properly classified based on photometric data only.
Measurements of their surface temperature or only a rough determination
of the spectral type from low-resolution spectra should provide discrimination
between these types of variables. The known $\beta$ Cep type stars are of
spectral types O9--B5 (based on data in Stankov and Handler 2005, Pigulski
and Pojma\'nski 2008ab) or surface temperatures in the range 16~000--36~000 K.
The $\delta$ Sct type stars are of spectral types A0--F9 or surface
temperatures in the range 6000--9000 K (based on catalog in Rodr\'iguez
\etal 2000).

We observed a dozen of the sixty unclassified pulsators, mostly the
brightest ones. Their light curves can be seen in Fig.~1. Low-resolution
spectra of these objects are shown in Figs.~2 and 3, in which we also
present spectra of a {\it bona fide} $\delta$ Sct type star OGLE-GD-DSCT-0011
(Pietrukowicz \etal 2013b) and $\beta$ Cep type star ASAS111308-6106.8
(Pojma\'nski 2002, Pigulski and Pojma\'nski 2008b). Spectral types of
the stars are between A5 and F2, thus all of these pulsating variables are
of $\delta$ Sct type. Since their period ratios differ from those in radially
pulsating $\delta$ Sct stars, it is very likely that many of the observed
periodicities come from non-radial modes. We give new names, refered
to the variability type of these stars, in the form OGLE-GD-DSCT-NNNN,
where NNNN is a four digit consecutive number starting from 0059 (see Table~3).

\begin{figure}[htb!]
\centerline{\includegraphics[angle=0,width=110mm]{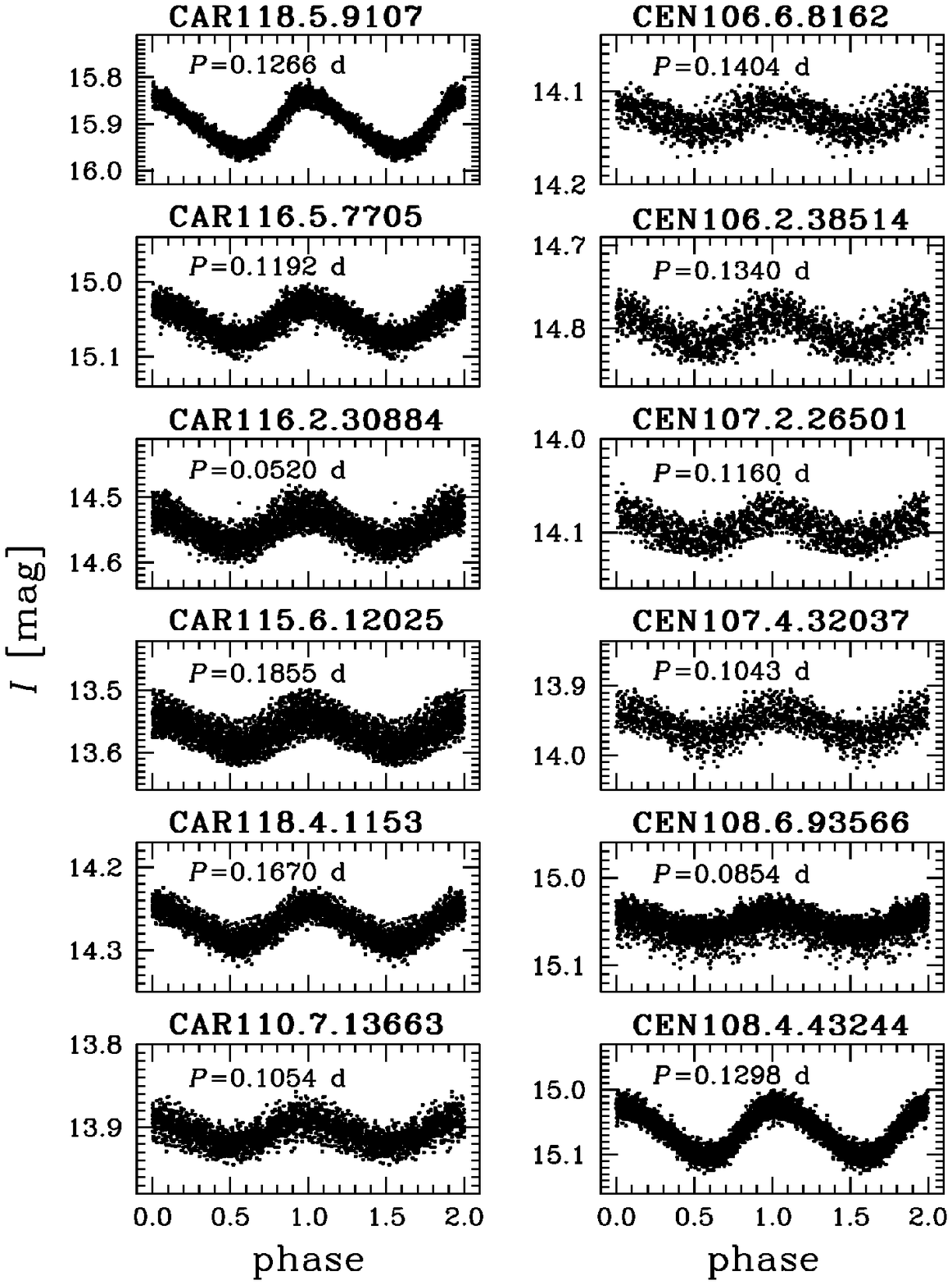}}
\FigCap{Light curves of twelve previously unclassified short-period
multi-mode pulsating stars phased with the main period. Spectra of
the stars in the left column are presented in Fig.~2, while in the right
one in Fig.~3.}
\end{figure}

\begin{figure}[htb!]
\centerline{\includegraphics[angle=0,width=130mm]{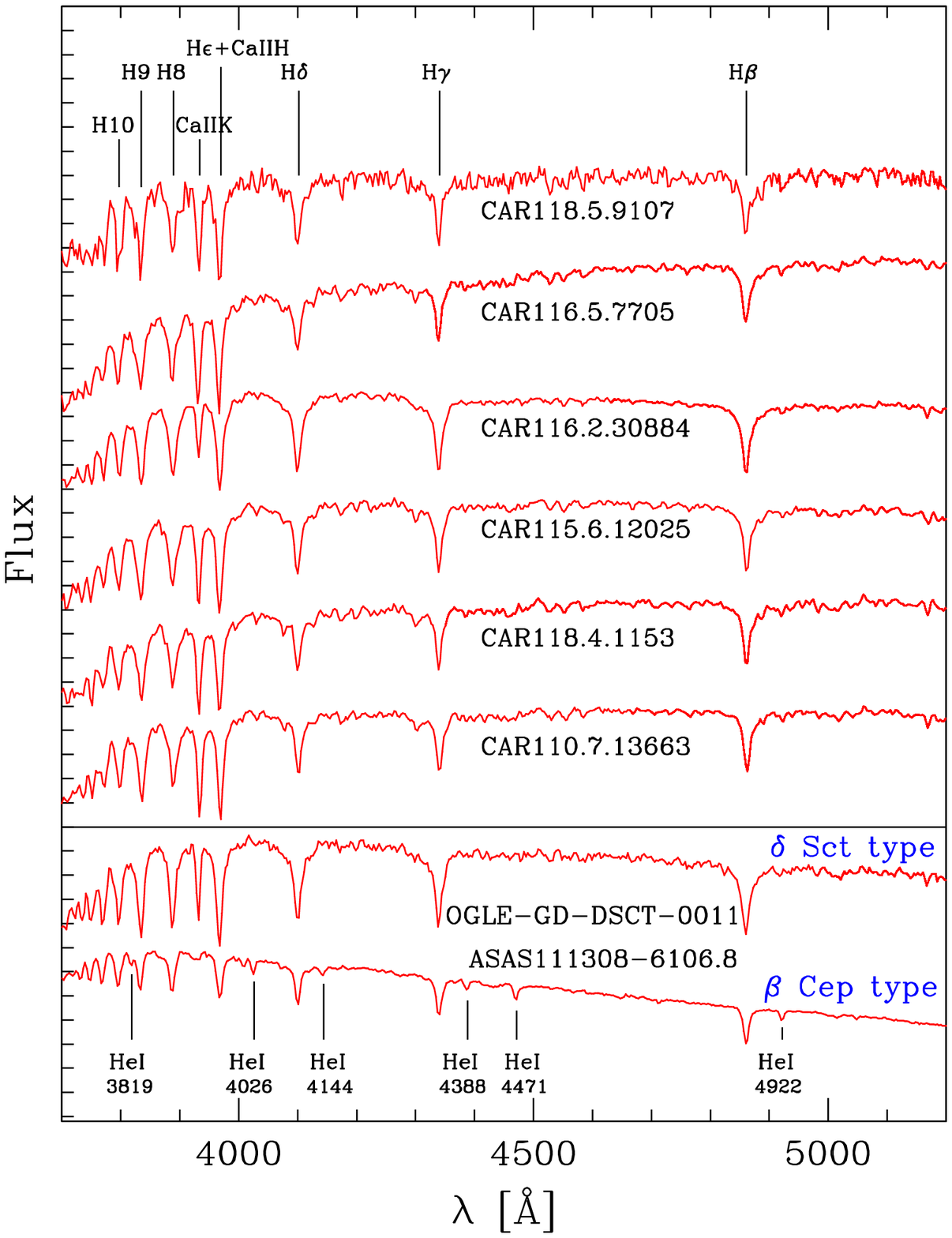}}
\FigCap{Spectra of six previously unclassified short-period multi-mode
pulsating stars detected in the OGLE-III disk fields in comparison
to the spectra of stars known as the $\delta$ Sct (OGLE-GD-DSCT-0011)
and $\beta$ Cep type (ASAS111308-6106.8). All six stars have spectra
very similar to that of OGLE-GD-DSCT-0011 and therefore all of the
variables are of $\delta$ Sct type.}
\end{figure}

\begin{figure}[htb!]
\centerline{\includegraphics[angle=0,width=130mm]{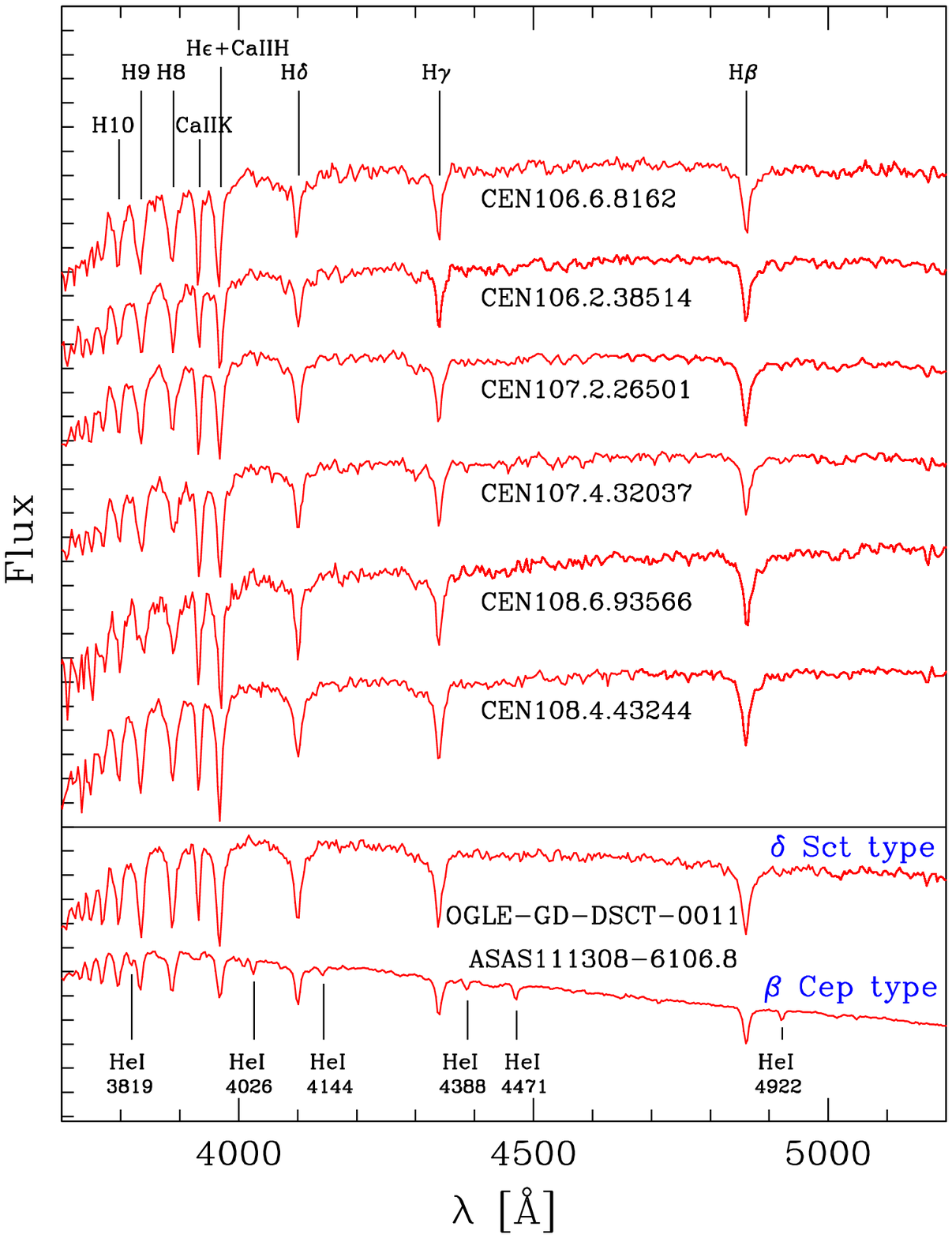}}
\FigCap{Spectra of another six previously unclassified short-period multi-mode
pulsators from OGLE-III in comparison to the spectra of a $\delta$ Sct type
and $\beta$ Cep type variables. The spectra clearly indicate that the six
variables are also of $\delta$ Sct type.}
\end{figure}

We also observed four other, likely pulsating stars, which spectra are
presented in Fig.~4. One of them is OGLE variable MUS101.3.34906
with a short period of 0.14610250(1)~d $\approx$~3.50646~h and an unusual
looking light curve. Period analysis reveals another two periodicities,
but with suspicious period ratios: $P_2=0.800083 P_1$ and $P_3=0.666780 P_1$
(see light curves in Fig.~5). The observed period ratios would indicate
a $\delta$ Sct type star pulsating in the first-, second-, and third-overtone
modes. The obtained A type spectrum confirms such identification.
We name this object OGLE-GD-DSCT-0071.

\begin{figure}[htb!]
\centerline{\includegraphics[angle=0,width=130mm]{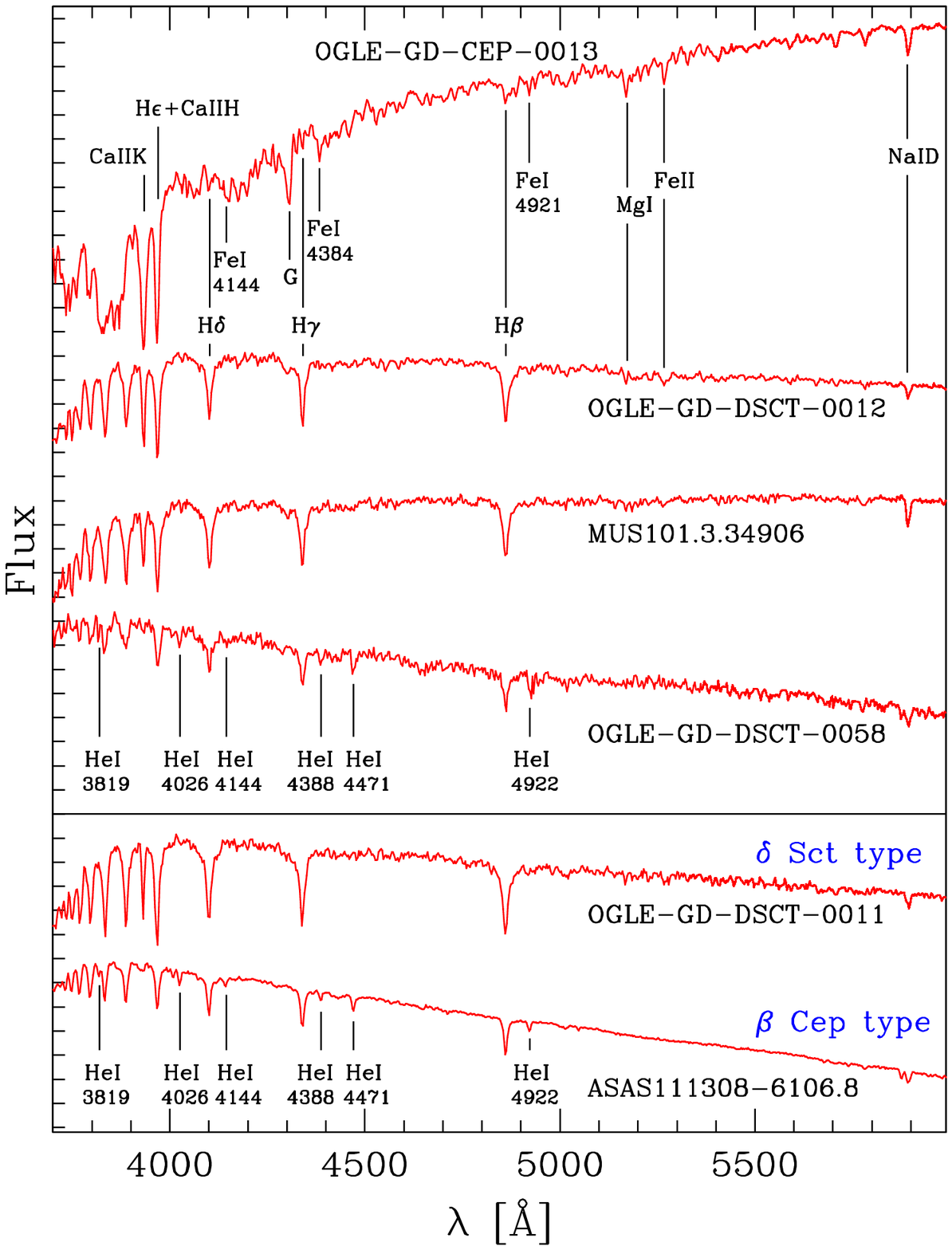}}
\FigCap{Spectra of four peculiar OGLE variables compared to the spectra
of stars known as the $\delta$ Sct and $\beta$ Cep type.}
\end{figure}

\begin{figure}[htb!]
\centerline{\includegraphics[angle=0,width=130mm]{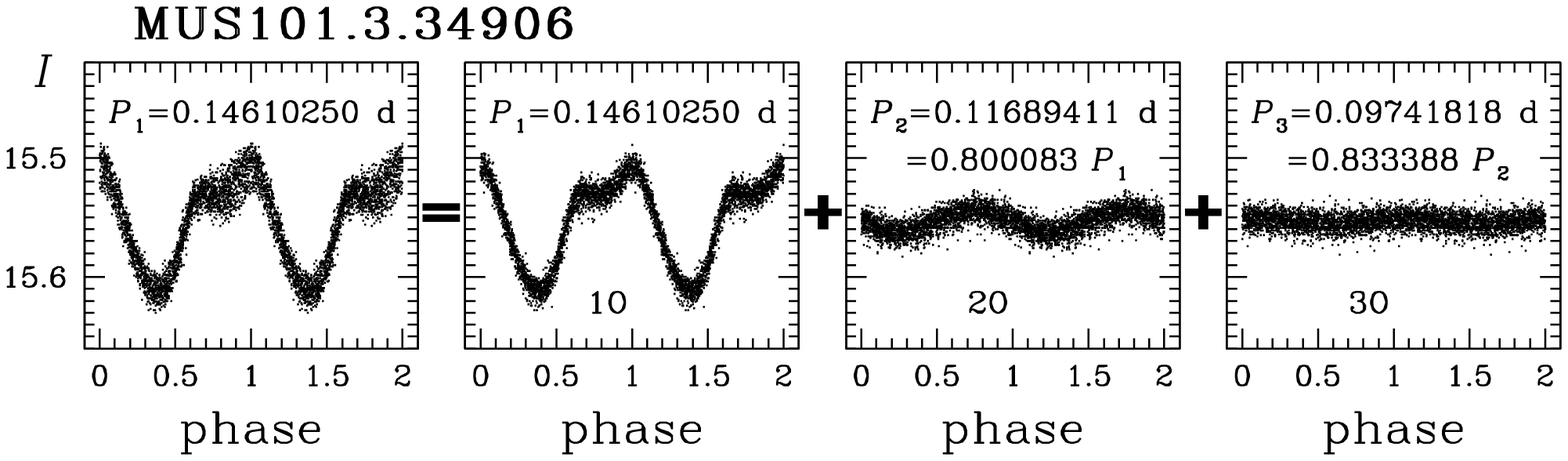}}
\FigCap{Decomposition of the light curve of object MUS101.3.34906,
confirmed to be a triple-mode $\delta$ Sct type star.}
\end{figure}

Another investigated target, a double-mode star OGLE-GD-DSCT-0012
could not be reproduced by models in the Petersen diagrams, shown
in Fig.~11 in Pietrukowicz \etal (2013b). The obtained spectrum confirms
that this is indeed a $\delta$ Sct type variable. Probably, one of the
modes is non-radial.

In Fig.~4, we present a spectrum of OGLE-GD-CEP-0013. This object with
the period of 5.2436~d seems to have a peculiar light curve if one
looks at the sequence of light curves of variables classified as classical
Cepheids and ordered by period in Fig.~1 in Pietrukowicz \etal (2013b).
The rising part lasts for about half of the period, in comparison
to about 20\% of $P=5.3664$~d in the case of the prototype star $\delta$~Cep
itself and about 35\% of $P=4.8609$~d in the first-overtone star V335~Pup
(see. Fig.~6). The obtained spectrum shows a continuum strongly
inclined to the red with weak Balmer series in absorption and many metal lines,
such as the Ca{\small II} K+H doublet ($\lambda\lambda$ 3934+3968 \AA),
the Mg{\small I} triplet ($\lambda\lambda$ 5167, 5173, 5184 \AA),
Fe{\small II} $\lambda$ 5267 \AA, the Na {\small I} D doublet
($\lambda\lambda$ 5890, 5896 \AA), and the G band formed of Fe, Ti, Ca lines
around 4307 \AA. Such a spectrum is characteristic for a metal-rich star
around G0 type. Classical Cepheids cover a wide range of spectral types
between F5 and K2. The observed shape of the light curve with an $I$-band
amplitude of 0.29~mag in a G0 type star can only be explained as pulsations.
OGLE-GD-CEP-0013 pulsates in the first-overtone mode rather than in the
fundamental mode.

\begin{figure}[htb!]
\centerline{\includegraphics[angle=0,width=80mm]{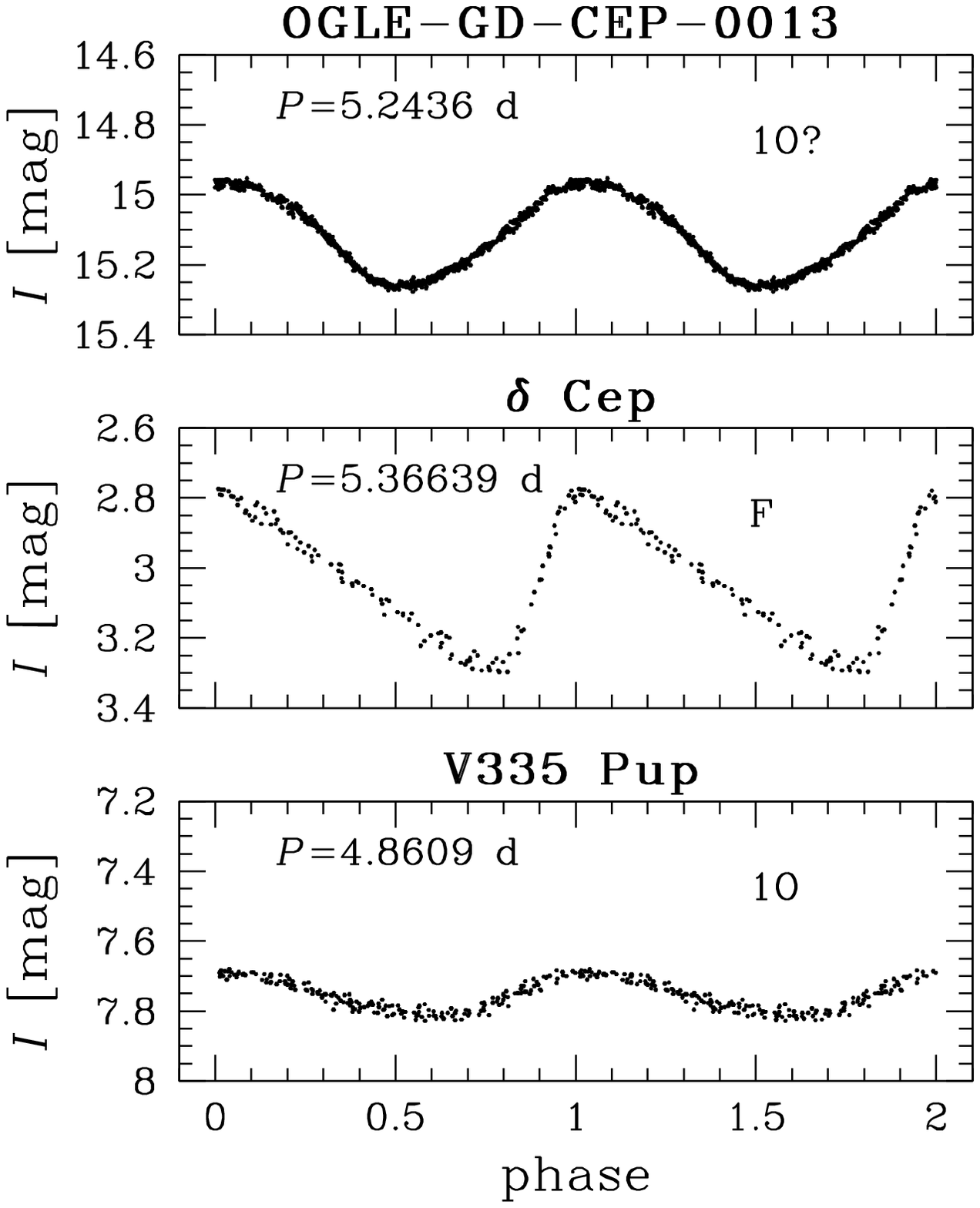}}
\FigCap{Comparison of the $I$-band light curve of object
OGLE-GD-CEP-0013 with light curves of two known classical
Cepheids of similar periods: the prototype star $\delta$~Cep pulsating
in the fundamental mode and V335~Pup pulsating in the first-overtone
mode. The magnitude range is the same in all panels. It seems that
OGLE-GD-CEP-0013 is a first-overtone pulsator. The data for $\delta$~Cep
come from the AAVSO observations, while for V335~Pup from the ASAS
project (Pojma\'nski 2002).}
\end{figure}


\Subsection{Mysterious Object OGLE-GD-DSCT-0058}

In Fig.~4, we also present the low-resolution spectrum of object
OGLE-GD-DSCT-0058. In Pietrukowicz \etal (2013b), it was classified
as a $\delta$ Sct type star with a very short period of 0.01962154(1)~d
= 28.255018(1) min. The power spectrum revealed two additional peaks
equally distant from the dominant mode in the frequency space,
$\pm1.9028$~d$^{-1}$ from 50.9644~d$^{-1}$, and of a similar power
($I$-band amplitudes $\approx0.025$~mag \vs the dominant mode of 0.231~mag). 
The three peaks were interpreted as a dipolar triplet, in which the
dominant peak is due to a radial mode and the two side oscillations are
dipole modes split by stellar rotation. Fig.~7 shows the overall and
decomposed light curve of this object.

\begin{figure}[htb!]
\centerline{\includegraphics[angle=0,width=130mm]{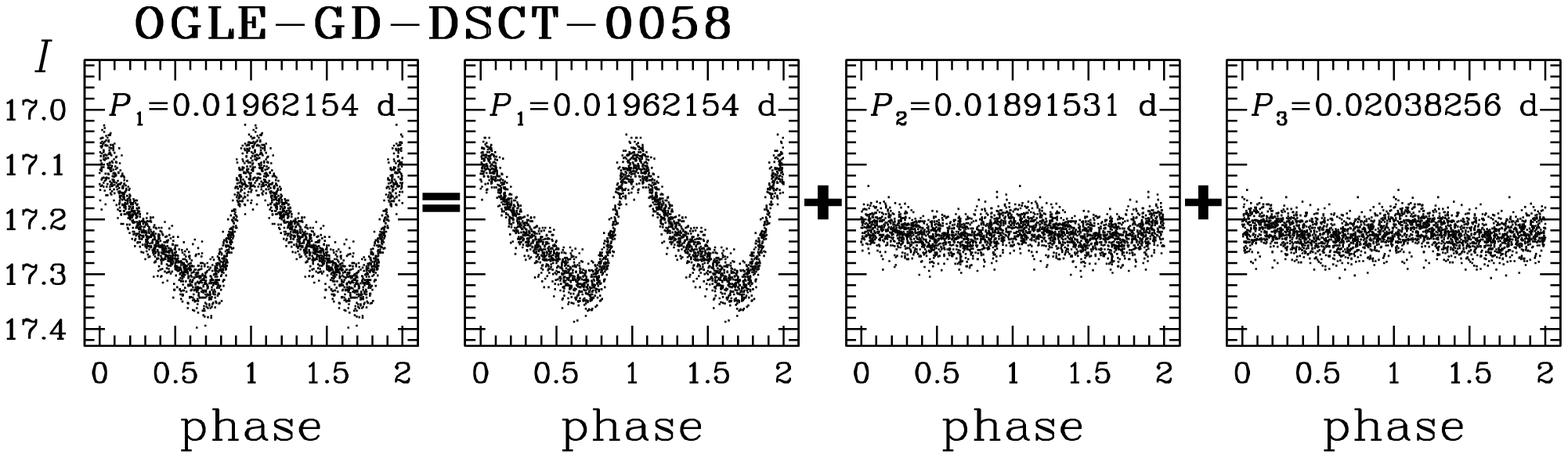}}
\FigCap{Decomposition of the light curve of OGLE-GD-DSCT-0058.}
\end{figure}

Surprisingly, the obtained low-resolution spectrum is characterized by a
strong continuum rising to the blue. Superimposed are relatively weak
hydrogen lines (when compared to typical $\delta$ Sct spectra) as well
as helium ones, thus indicating a high temperature for the object.
A quick literature search allowed us to find some similarities between
the spectrum of OGLE-GD-DSCT-0058 and those of moderately helium-enriched
hot subdwarf stars of B and O type (sdOB, Vennes \etal 2007, Drilling \etal 2013).
This is why we tentatively fitted the Balmer and helium lines observed
with a grid of TLUSTY non-local thermodynamic equilibrium (NLTE) model
atmospheres suited for hot subdwarf stars (see Brassard \etal 2010,
Latour \etal 2014). The model grid covers a rather large range of
parameters: 20~000~K \textless\ $T_{\rm eff}$ \textless\ 50~000~K,
4.6 \textless\ log~$g$ \textless\ 6.4, and finally
$-$4.0 \textless\ log~$N$({\rm He})/$N$({\rm H}) \textless\ 0.0.
The metallicity used in the models is one appropriate for typical hot
subdwarfs, as measured by Blanchette \etal (2008): solar abundances for
N, S and Fe, and one tenth solar for C, O, and Si. The best fit solution
is shown in Fig.~8, note that the uncertainties are the formal ones given
by the $\chi^2$ minimization procedure. The high temperature found
($T_{\rm eff} \approx 33~000$~K, corresponding to type O9) confirms that
this object cannot be a single $\delta$ Sct type star. The high surface gravity
of log~$g=5.3\pm0.2$, which is much higher than values for high-mass
main sequence stars (log $g$ between 4.2 and 4.3, according to models
in Pamyatnykh 1999), and the high amplitude at the extremely short
period rule out the possibility that this is a $\beta$ Cep type star.

\begin{figure}[htb!]
\centerline{\includegraphics[angle=-90,width=130mm]{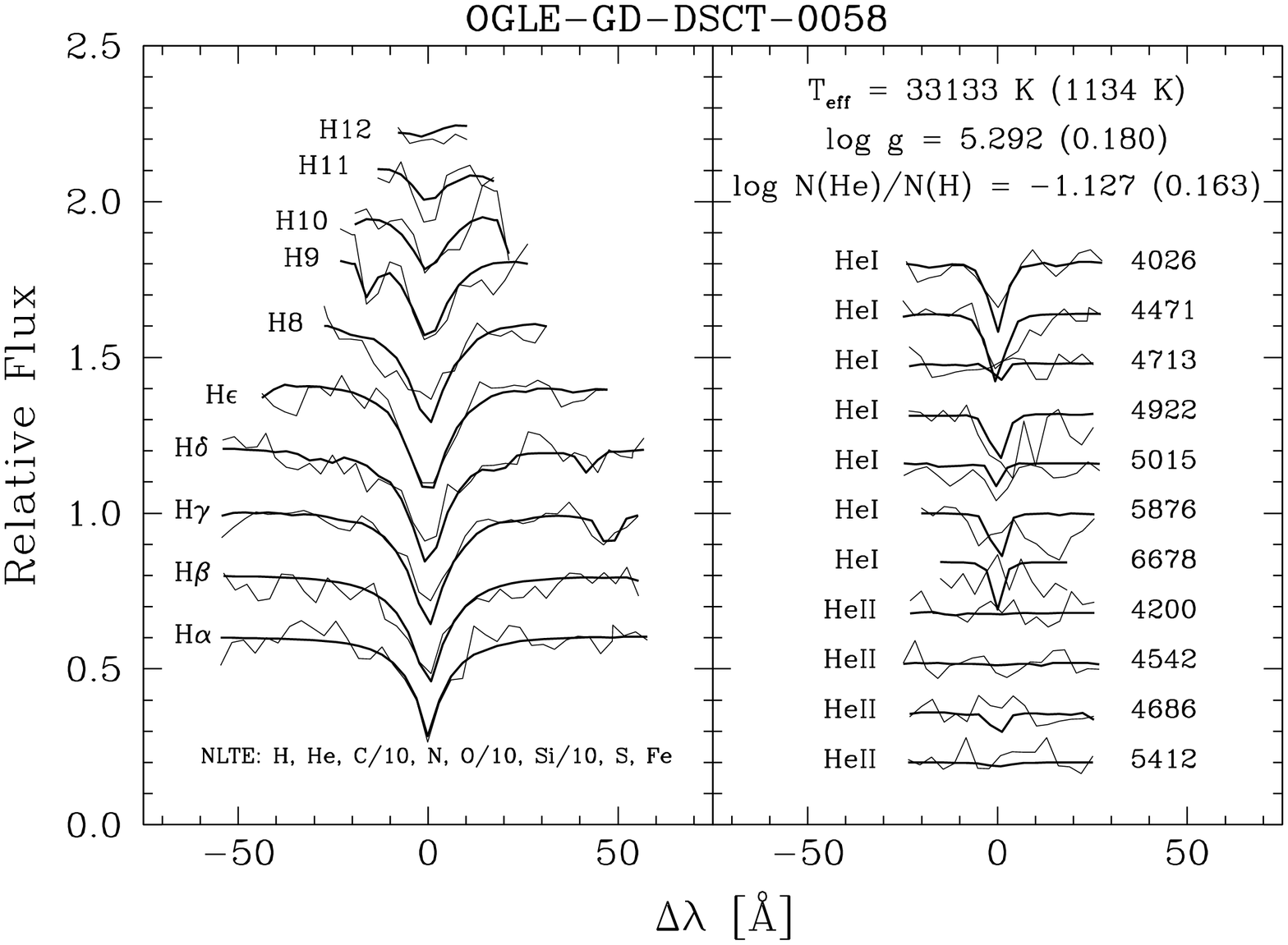}}
\FigCap{Best fit (thick line) to the hydrogen and helium lines of
observed spectrum (thin line) of OGLE-GD-DSCT-0058.
For each spectral line, the continuum is normalized to 1.0
and shifted by 0.2 in flux units.}
\end{figure}

The interpretation of OGLE-GD-DSCT-0058 as a pulsating hot subdwarf
does not agree with predicted and observed pulsation properties
for this class of objects (Charpinet \etal 2009, Heber \etal 2009).
The rapid $p$-mode sdOB pulsators (of V361 Hya type) having
$T_{\rm eff}$ from 28~000~K to 35~000~K and log~$g$ between 5.2
and 6.1 oscillate with periods between about 1.0~min and 10~min
and amplitudes rarely exceeding 0.01~mag in $V$ (0.12~mag in the
most extreme case of star Balloon 090100001, Oreiro \etal 2004,
Baran \etal 2009). The cooler $g$-mode pulsators (of V1093 Her type)
with $T_{\rm eff}$ between 23~000~K and 30~000~K and log~$g$ around 5.4
have periods ranging from about 45~min to 120~min and amplitudes
lower than $p$-mode stars (of millimagnitudes).

The obtained surface gravity for OGLE-GD-DSCT-0058 is far too low
for a white dwarf, while the temperature far too low for a planetary
nebula nucleus. Moreover, pulsating objects from the above classes show
much lower amplitudes. The shape of the light curve of OGLE-GD-DSCT-0058
resembles those of high-amplitude fundamental-mode pulsators in the main
instability strip. There is a possibility that OGLE-GD-DSCT-0058
is not a single object. We cannot rule out an option that the observed
light variations are not due to pulsations. High-resolution spectroscopic
observations should help to reveal the true nature of this object.


\Subsection{Variables with Irregular Fluctuations}

Three of the observed stars are characterized by irregular light variations
on time scales from hours to days and amplitudes reaching 2.0 mag in $I$,
as it can be seen in Fig.~9. Based on the OGLE-III photometry star
CAR117.5.6157 might be classified as a Z Cam type dwarf nova.
This type of cataclysmic variables exhibit standstills during which
outbursts cease for days to years. The photometric behavior of the other
two stars, MUS100.3.59946 and BLG179.1.111115, resembles light changes
observed in hydrogen-deficient stars of DY Per and R CrB type.

In Fig.~10, we present the spectra of the three interesting
objects. They are very similar to each other. The spectra are characterized
by a flat or red continuum upon which Balmer series is superimposed.
Whereas the higher members of the series are in absorption, H$\alpha$ is
found in emission. Other recognizable features are absorption lines of
Ca{\small II}, Mg{\small I}, Fe{\small II}, Na{\small I} D, and the G band.
This kind of spectra are typical for young stellar objects (see examples
in Kamath \etal 2014). The spectral types between A5 and G3 indicate
that these stars are low-mass objects ($<3\MS$) of T Tau type. We searched
for the presence of lithium, the line Li{\small I} $\lambda$ 6708 \AA.
Lithium is expected to be abundant in the matter from which the star
was formed, but it can be depleted already in the pre-main sequence phase.
We were not able to detect this element in our low-resolution spectra.

\begin{figure}[htb!]
\centerline{\includegraphics[angle=0,width=130mm]{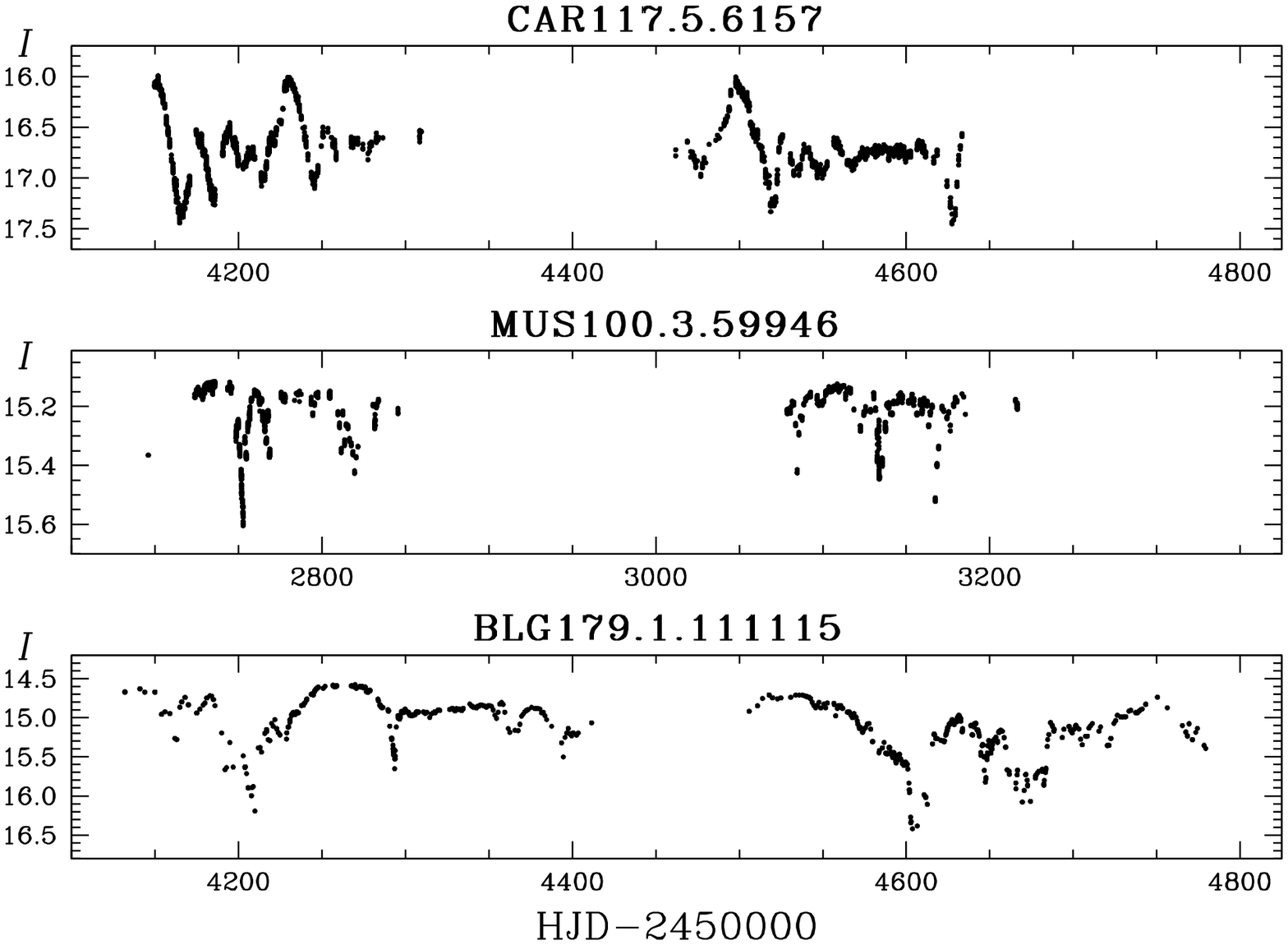}}
\FigCap{Three variables with irregular fluctuations. The light variations
in the first star are similar to those observed in Z Cam type dwarf novae,
while in the other two objects to hydrogen-deficient stars of DY Per type.
Their spectra, presented in Fig.~10., indicate that all of them
are young stellar objects.}
\end{figure}

\begin{figure}[htb!]
\centerline{\includegraphics[angle=90,width=120mm]{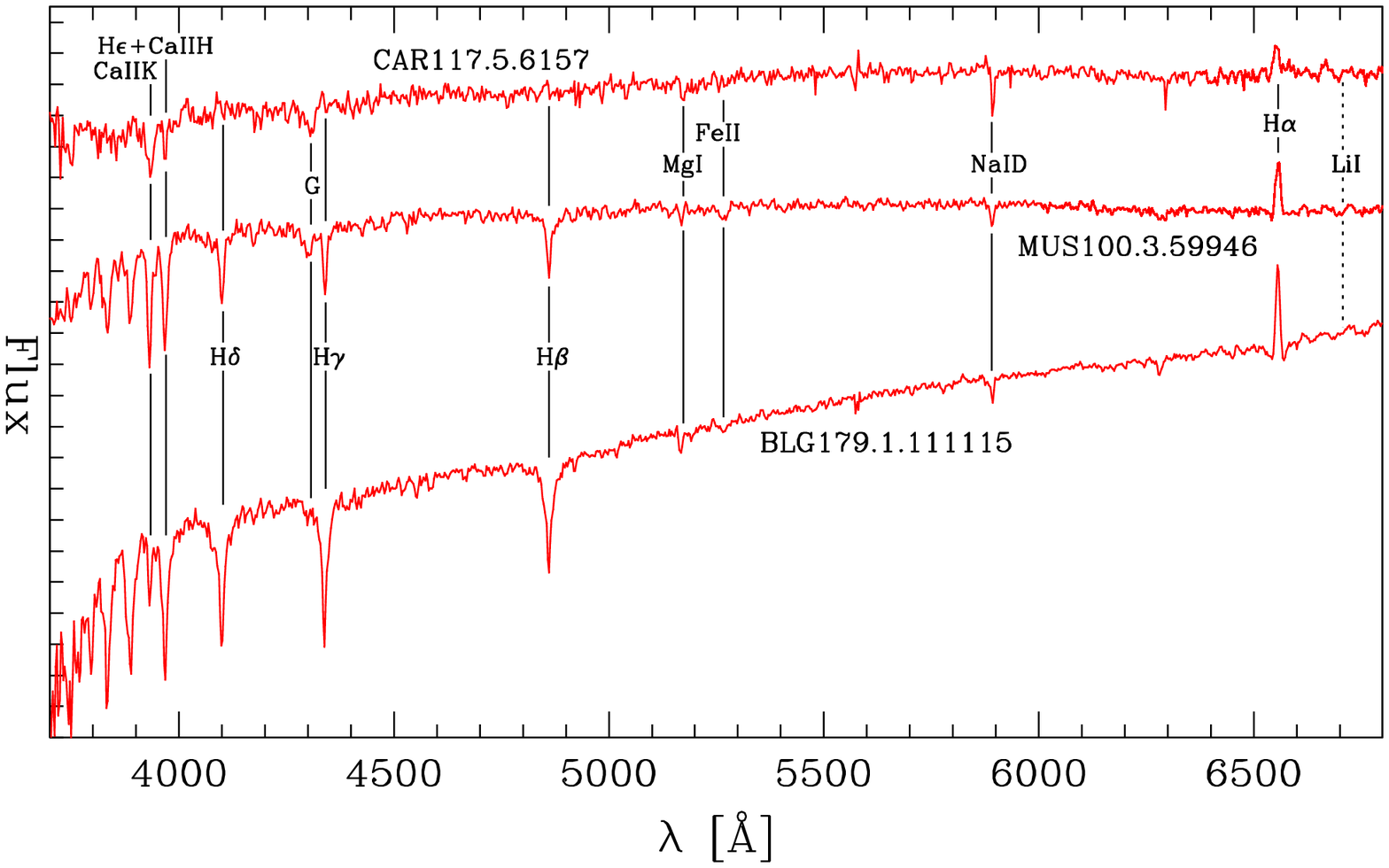}}
\FigCap{Spectra of the three variables with irregular fluctuations.
All stars show deep Ca{\scriptsize II} HK and Na{\scriptsize I} D lines,
and H$\alpha$ in emission. The G band is particularly prominent in CAR117.5.6157.
All these objects are of T Tau type.}
\end{figure}


\Subsection{Periodic Variables with Brightenings}

We also took spectra of five periodic objects displaying brightenings or
waves with $I$-band amplitudes up to 0.2~mag in their phased light curves
(Fig.~11). The objects show a single or double wave over the period,
which is between 1.5~d and 6.6~d. In the case of CAR106.7.14162, the shape
and amplitude of the light variations are not stable. In the other four
stars, the waves seem to be stable over the years. Some of them last for
more than half of the period. The spectra of these stars, presented in Fig.~12,
show strong hydrogen absorption lines characteristic for spectral type A
and also enhanced lines of elements such as silicon.
All five objects are very likely chemically-peculiar Ap type stars
with a relatively strong global magnetic field which axis is inclined
to the rotation axis, as it is described in the oblique-rotator model
(Stibbs 1950, Preston 1967). The observed single or double wave is
a result of changing aspect of their spotted surface with the rotation
period (\eg St\c{e}pie\'n 1968, Catalano and Leone 1991, Manfroid and Renson
(1994), Bychkov \etal 2005). This kind of variables are classified as
$\alpha^2$ CVn type stars.

\begin{figure}[htb!]
\centerline{\includegraphics[angle=0,width=130mm]{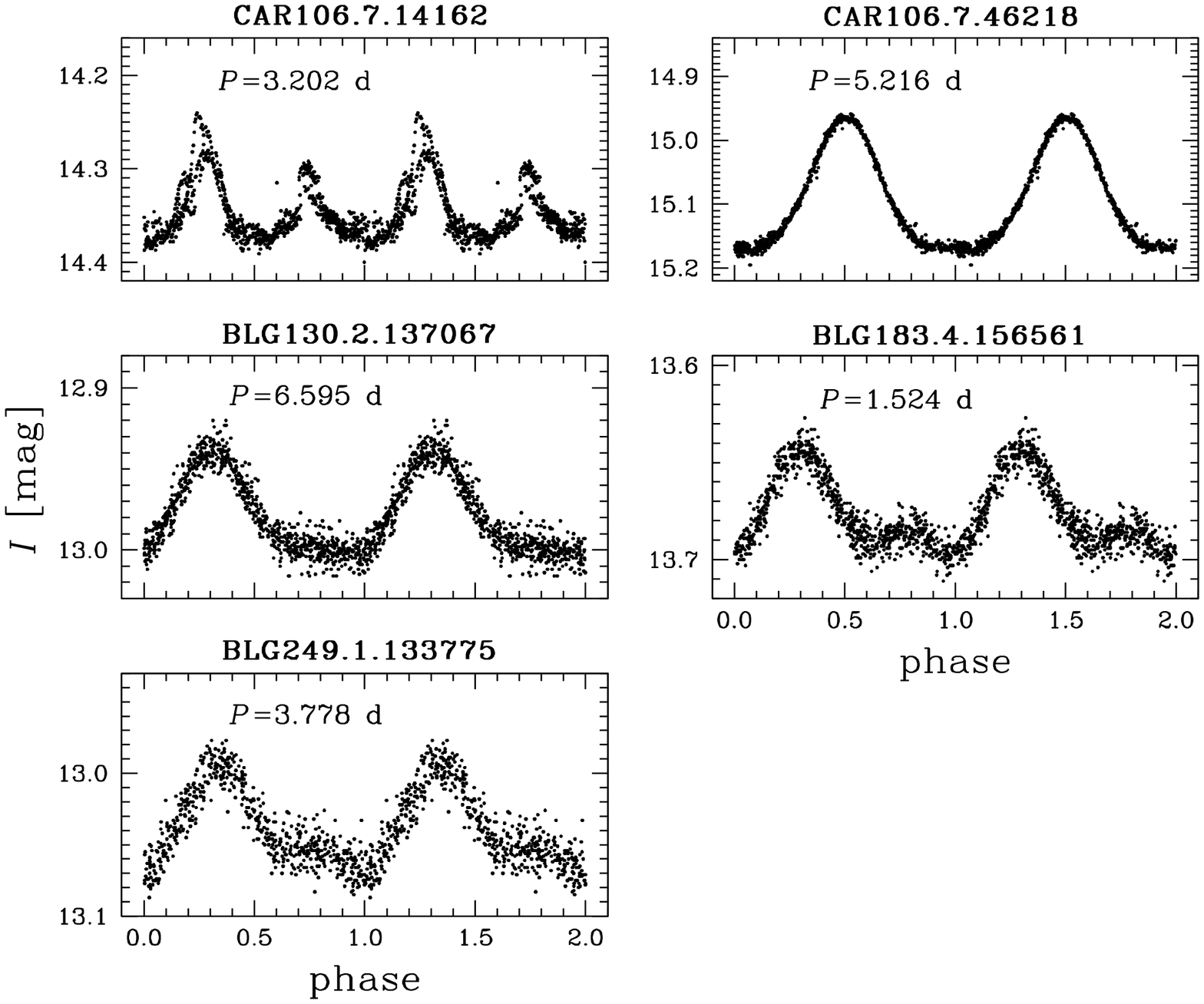}}
\FigCap{Phased light curves of five OGLE stars showing periodic
brightenings or waves.}
\end{figure}

\begin{figure}[htb!]
\centerline{\includegraphics[angle=0,width=130mm]{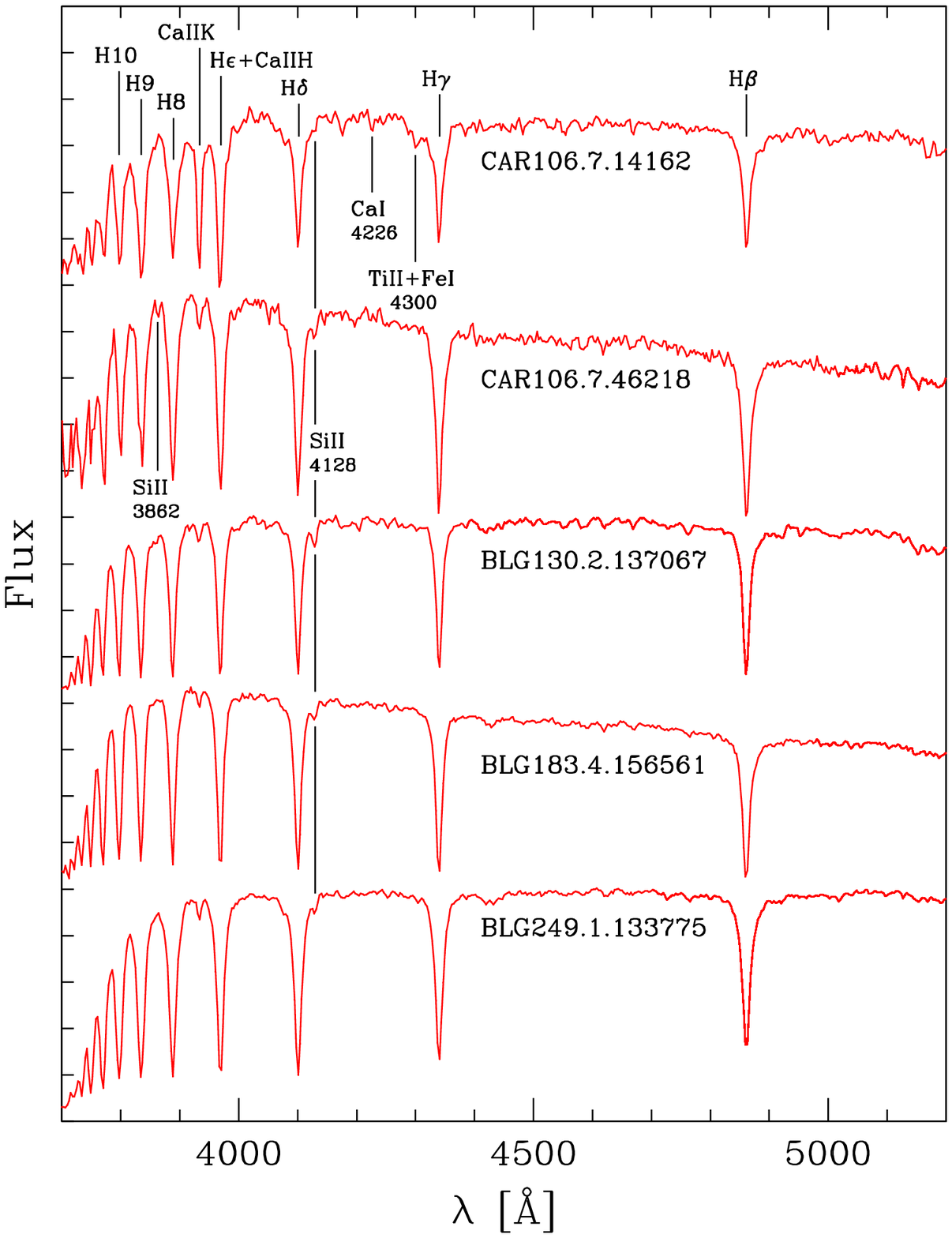}}
\FigCap{Spectra of the stars with periodic brightenings. These are very
likely Ap stars of Si type.}
\end{figure}


\Subsection{Summary}

We summarize the results of our spectroscopic follow-up in Table~3.
For all but one object the obtained low-resolution spectra have allowed
us to make a definitive classification. In the case of OGLE-GD-DSCT-0058,
the spectrum shows that it cannot be a single $\delta$ Sct type star.
The observed properties of this object do not fit to any kind of known
pulsating variables. We hope that high-resolution spectra taken over
different phases will help to answer the question on the true nature
of this enigmatic object.

\begin{table}[htb!]
\centering \caption{\small Final classification of the target OGLE variables}
\medskip
{\small
\begin{tabular}{lccc}
\hline
ID or name        & Sp. type & Var. type         & New name \\
\hline
CAR118.5.9107     & F0 & $\delta$ Sct & OGLE-GD-DSCT-0059 \\
CAR116.5.7705     & F0 & $\delta$ Sct & OGLE-GD-DSCT-0060 \\
CAR116.2.30884    & A5 & $\delta$ Sct & OGLE-GD-DSCT-0061 \\
CAR115.6.12025    & F0 & $\delta$ Sct & OGLE-GD-DSCT-0062 \\
CAR118.4.1153     & F0 & $\delta$ Sct & OGLE-GD-DSCT-0063 \\
CAR110.7.13663    & F0 & $\delta$ Sct & OGLE-GD-DSCT-0064 \\
CEN106.6.8162     & F0 & $\delta$ Sct & OGLE-GD-DSCT-0065 \\
CEN106.2.38514    & A7 & $\delta$ Sct & OGLE-GD-DSCT-0066 \\
CEN107.2.26501    & F0 & $\delta$ Sct & OGLE-GD-DSCT-0067 \\
CEN107.4.32037    & F2 & $\delta$ Sct & OGLE-GD-DSCT-0068 \\
CEN108.6.93566    & F0 & $\delta$ Sct & OGLE-GD-DSCT-0069 \\
CEN108.4.43244    & F0 & $\delta$ Sct & OGLE-GD-DSCT-0070 \\
MUS101.3.34906    & A7 & $\delta$ Sct & OGLE-GD-DSCT-0071 \\
OGLE-GD-DSCT-0012 & F0 & $\delta$ Sct & - \\
OGLE-GD-DSCT-0058 & O9 & ? & - \\
OGLE-GD-CEP-0013  & G0 & $\delta$ Cep & - \\
\hline
CAR117.5.6157     & G3 & T Tau & OGLE-GD-YSO-0001 \\
MUS100.3.59946    & F5 & T Tau & OGLE-GD-YSO-0002 \\
BLG179.1.111115   & A5 & T Tau & OGLE-BLG-YSO-0001 \\
\hline
CAR106.7.14162    & A7p & $\alpha^2$ CVn & OGLE-GD-ACV-001 \\
CAR106.7.46218    & A0p & $\alpha^2$ CVn & OGLE-GD-ACV-002 \\
BLG130.2.137067   & A0p & $\alpha^2$ CVn & OGLE-BLG-ACV-001 \\
BLG183.4.156561   & A0p & $\alpha^2$ CVn & OGLE-BLG-ACV-002 \\
BLG249.1.133775   & A0p & $\alpha^2$ CVn & OGLE-BLG-ACV-003 \\
\hline
\end{tabular}}
\end{table}


\Acknow{
It is a pleasure to acknowledge discussions with Dr. W. Dziembowski
and Dr. K. St\c{e}pie\'n. We thank Dr. G. Pojma\'nski for providing
ASAS data. In this work, we have also used observations from the
AAVSO International Database contributed by observers worldwide.
This work has been supported by the Polish Ministry of Sciences
and Higher Education grants No. IP2012 005672 under the Iuventus Plus
program to PP and No. IdP2012 000162 under the Ideas Plus program
to IS. The OGLE project has received funding from the European
Research Council under the European Community$'$s Seventh Framework
Programme (FP7/2007-2013)/ERC grant agreement No. 246678 to AU,
PI of the project. ML acknowledges funding by the Deutsches
Zentrum f\"ur Luft- und Raumfahrt (grant 50 OR 1315).
CG acknowledges full financial support from the postdoctoral
fellowship program PNPD/CAPES-Brasil.}


\end{document}